\documentclass[intlimits,twoside,a4paper]{article}
\usepackage[cp1251]{inputenc}

\usepackage[eqsecnum]{cmpj3}

\DeclareFontEncoding{LGR}{}{}
\DeclareRobustCommand{\greektext}{%
	  \fontencoding{LGR}\selectfont\def\encodingdefault{LGR}}
\DeclareRobustCommand{\textgreek}[1]{\leavevmode{\greektext #1}}

\newcommand{\lyxmathsym}[1]{\ifmmode\begingroup\def\b@ld{bold}
	  \text{\ifx\math@version\b@ld\bfseries\fi#1}\endgroup\else#1\fi}

\providecommand{\tabularnewline}{\\}

\issue{2025}{28}{4}{43801}

\title[Investigation of density of states and charge carrier mobility in amorphous semiconductors via time-of-flight]
{Investigation of density of states and charge carrier mobility in amorphous semiconductors via time-of-flight photocurrent analysis}
\author[F. Serdouk, A. Boumali, M. L. Benkhedir, Y. Goutal]{
F. Serdouk\orcid{0009-0001-7274-9737}\thanks{fadila.serdouk@univ-tebessa.dz}, 
A. Boumali\orcid{0000-0003-2552-0427}\thanks{Corresponding author: \email{boumali.abdelmalek@gmail.com}.},
M. L. Benkhedir\orcid{0000-0001-8375-0998}\thanks{benkhedir@gmail.com},
Y. Goutal\orcid{0000-0001-5747-9081}\thanks{yazid.gouttel@univ-tebessa.dz}
}

\address{
Laboratory of Theoretical and Applied Physics, Echahid Cheikh Larbi Tebessi University, Tebessa, Algeria}


\date{Received February 11, 2025, in final form September 25, 2025}
\begin{document}
\maketitle
\begin{abstract}
{The present study examines the electronic transport
characteristics of amorphous semiconductors through TOF
 measurements and numerical simulations. The primary objective
is to determine the DOS in amorphous selenium
(a-Se) and to assess the temperature and electric field dependence
of the hole mobility. A comprehensive investigation of localized states
within the mobility gap is performed using Laplace transform analysis
of ToF photocurrent transients, combined with the multiple trapping
model. This approach enables accurate reconstruction of the DOS across
a wide temperature range, allowing clear identification of shallow
and deep trap levels and revealing thermally activated transport mechanisms.
Simulated ToF currents are also used to evaluate the hole drift mobility
under various thermal and field conditions. Activation energies are
extracted from Arrhenius plots of the mobility data. The results support
a physically consistent description of the electronic structure in
a-Se and validate the applicability of Laplace-based techniques for
probing charge transport in disordered semiconductors.}
\printkeywords 
\end{abstract}

\section{Introduction}

Amorphous semiconductors, especially amorphous selenium (a-Se), have
attracted considerable interest owing to their distinctive electrical
characteristics and practical uses in devices like photoreceptors
and X-ray detectors. In contrast to crystalline semiconductors, a-Se
exhibits an absence of long-range order, leading to a wide distribution
of localized states inside its mobility gap. Comprehending the electronic
transport characteristics of a-Se, including the 
DOS and charge carrier mobility is essential for enhancing its
efficacy in electrical and optoelectronic devices. The movement of
charge within a dielectric substance is crucial in numerous electrical
and photographic systems. A comprehensive understanding of process
dynamics can yield significant insights into the electrical
structure of the material. The temporal rate of charge displacement and the efficiency
of their creation by light or energetic electrons have been investigated
experimentally using a ToF technique (for further details
on this technique, see to \cite{spear1960,schmidlin1977,noolandi1977,schmidlin1980,Kasner1983}).

The assessment of the DOS in amorphous semiconductors
yields significant understanding of the energy distribution of localized
states, which serve as trapping foci for charge carriers. These states
profoundly affect the transport parameters and are essential for comprehending
charge recombination, trapping, and detrapping procedures. This study
aims at characterizing the DOS in amorphous selenium
(a-Se) utilizing modern methodologies, including Laplace transform
analysis of transient photocurrent (TPC) data, alongside ToF
(ToF) measurements to evaluate the hole mobility. The ToF approach is
among the most prevalent methods for quantifying the charge carrier mobility
in disordered materials such as a-Se. In a standard ToF experiment,
a brief pulse of light produces electron-hole pairs within the material.
An applied electric field induces a drift of charge carriers, and
their transit time is measured as they traverse the material.

The mobility $\mu$ of the charge carriers is determined using the
equation $\mu={L}/ F\cdot t_{\mathrm{tr}}$, where $L$ represents
the sample thickness, $F$ denotes the applied electric field, and
$t_{\mathrm{tr}}$ signifies the transit time. The ToF approach delivers
direct data regarding charge transport characteristics, elucidating the
aspects of carrier movement, trap distribution, and the characteristics
of localized states. The ToF approach is especially advantageous for
amorphous semiconductors as it differentiates between dispersive and
non-dispersive transport. In dispersive transport, carriers undergo
considerable trapping, resulting in a spreading of the current pulse,
whereas in nondispersive transport, carriers display uniform motion
with a precise transit time, commonly seen in materials with few traps
or isolated states. Consequently, the duration-of-Flight ToF measuring
technique is an essential instrument for assessing the transit duration
of mobile entities and is extensively used to investigate the charge carrier
transport characteristics in diverse condensed materials, encompassing
both inorganic and organic semiconductors. The resultant transient
current elucidates the transit duration and offers insights into dynamic
processes during the charge transport through the sample, including carrier
trapping and release in trap states, carrier diffusion, and percolation
(for further details, see to \cite{serdouk2015,serdouk2020,serdouk2023,goutal2024}).

Schmidlin and Noolandi \cite{schmidlin1977,noolandi1977} proposed
the multiple-trapping model (MTM) to elucidate the dynamic behavior
of charge carriers in a sample through traditional sets of linked
kinetic equations governing trapping and heat release rates. This
method can precisely reproduce the experimental current signal $I(t)$
when appropriate parameters for trapping and release rates are presumed.
The MTM has been extensively and effectively utilized to examine transient
current decays in both inorganic and diverse organic semiconductors.
These films comprise amorphous inorganic or organic semiconductors,
employing suitable assumptions for trapping and release parameters
represented by a continuous distribution of trap states, including
exponential $\re^{-E/k_{\text{B}}T}$, Gaussian, or other trap distributions.
This strategy is effective only when the selected trap distribution
shape appropriately represents the localized states that predominate
the trapping occurrences \cite{ohno2008,naito1994,naito1996}.

Naito et al. \cite{naito1994,naito1996,ogawa2000,nagase1998} introduced
a spectroscopic technique to derive localized-state distributions
by examining transient photo-current data acquired via transient photoconductivity
or the ToF method, employing the Laplace transform within a multiple-trapping
model and applying appropriate approximations. The Laplace transform-based
method presents multiple advantages compared to alternative techniques:
(i) it effectively extracts localized-state distributions for materials
exhibiting either nondispersive or dispersive transport; the analysis
is both computationally efficient and rapid, and (ii) it facilitates
the extraction of localized-state distributions from both the pre-
and post-monomolecular recombination phases in transient photoconductivity,
as well as from both pre- and post-transit time phases in ToF photo-current
transients. This facilitates an expanded measurement range of localized-state
distributions.

{A series of comprehensive and influential studies
significantly advanced the understanding of localized states in amorphous
selenium (a-Se). By employing multiple experimental techniques, these
works probed the electronic structure of a-Se and initiated a scientific
debate regarding the symmetry --- or asymmetry --- of the density of
localized states (DOS) near the band edges. In particular, by analyzing
both the pre-transit and post-transit regimes of ToF 
photocurrents, Benkhedir et al. \cite{benkhedir2004,Benkhedir2004a,Qamhieh2004,benkhedir2008,Benkhedir2010} demonstrated
the effectiveness of transient current analysis in revealing the distribution
of trap states on both sides of the bandgap.}

{Building on this foundational work, the present study
focuses specifically on pure stabilized amorphous selenium and aims
at refining the understanding of the density of localized states in
the conduction band tail. Our methodology introduces several key advancements
over the prior approaches. Firstly, we apply a Laplace transform to the
entire ToF photocurrent signal, without isolating specific temporal
regions (e.g., pre- or post-transit), thereby preserving the full
dynamics of charge transport. Secondly, our measurements span a wide
range of temperatures, allowing access to thermally activated transport
regimes and providing finer energy resolution in the extracted DOS.
This technique enables a direct, assumption-free reconstruction of
the conduction band DOS, in contrast to earlier methods based on iterative
fitting and predefined functional forms \cite{Koughia2005}.}

{This study provides a detailed temperature-resolved
extraction of the conduction band density of localized states (DOS)
in pure amorphous selenium using a Laplace transform-based analysis
of ToF currents --- an approach that offers enhanced resolution and
physical insight compared to the previous models. By addressing the limitations
of earlier methods and enriching the understanding of the electronic
structure in a-Se, this work strengthens the theoretical and experimental
foundations for modelling the charge transport in disordered semiconductors.}

{To assess the physical plausibility of the extracted
DOS and to explore a potential symmetry between conduction and valence
band tails, we rely on our previously developed model near the valence
band edge. This framework is employed to simulate ToF transients and
investigate how hole mobility evolves with temperature and electric
field. The corresponding numerical estimations are presented, along
with an analysis of the Poole--Frenkel effect and derived activation
energies. Unlike earlier assumptions \cite{Kasap2015}, which suggested
a monotonously increasing DOS extending up to 0.55\,eV above the
valence band edge, our results point to a more localized distribution
of states, offering a better alignment with the experimental transport
behavior.}

{By comparing the conduction band DOS extracted in
this work with the simulated behavior based on the valence band model,
we aim at providing a unified and physically consistent description
of the localized states in a-Se, and at reevaluating the assumption
of symmetry between both band tails under controlled experimental
conditions. In this context, the present work offers an alternative
extraction approach and provides a detailed examination of existing
DOS models, with the aim of supporting further progress in the understanding
and modelling the charge transport in disordered semiconductors.}

\section{Review of the laplace transform method for ToF photocurrent}

\subsection{Theory}

In ToF measurements, a thin film of amorphous semiconductors
is positioned between two electrodes, with at least one electrode
serving as a barrier to carrier injection. A brief pulse of highly
absorbed light stimulates a small layer of electron-hole pairs via
one of the electrodes. Depending on the polarity of the applied electric
field, either holes or electrons are induced into the bulk of the material.
The carriers subsequently arrive at the opposing electrode within
a specific transit time, $t_{\text{r}}$, which is recognized as either a
swift decrease in the transient photocurrent for nondispersive transport
or an inflection point in the double logarithmic representation of
the transient photocurrent for dispersive transport.

Several assumptions are established for the ToF Laplace Transform
(LT) analysis: (i) The transport of photogenerated carriers transpires
through a trap-controlled band transport mechanism; (ii) the mobile
carriers consist of either holes or electrons; (iii) all localized
states exhibit uniform capture cross sections; and (iv) the localized
states remain unsaturated, a condition commonly observed during ToF
transient photocurrent measurements under small-signal conditions.
The essential equations for the trap-controlled band transport mechanism
are articulated as \cite{schmidlin1977,noolandi1977}:
\begin{equation}
\frac{\partial p(x,t)}{\partial t}=-\sum_{i}\mu\frac{\partial p_{i}(x,t)}{\partial t}-\mu_{0}F\frac{\partial p(x,t)}{\partial x}+p_{0}\delta(t)\delta(x),
\label{eq:1}
\end{equation}
\begin{equation}
\frac{\partial p_{i}(x,t)}{\partial t}=\omega_{i}p(x,t)-\gamma_{i}p_{i}(x,t).
\label{eq:2}
\end{equation}
In this context, $x$ denotes the distance from the lighted surface,
$p(x,t)$ signifies the free carrier density at location $x$ and
time $t$, whereas $p_{i}(x,t)$ represents the trapped carrier density
at the $i$-th localized state. In this context, $p_{0}$ represents
the injected free carrier density per unit area, while $\omega_{i}=\sigma_{p}\nu_{\text{th}}g\left(E_{i}\right)\cdot\Delta E$
represents the capture rate constant at the $i$-th localized state,
$\gamma_{i}=\nu\exp\left(-E_{i}/k_{\text{B}}T\right)$ denotes the release rate
constant at the $i$-th localized state, $\sigma_{p}$ signifies the
capture cross section, $\nu_{\text{th}}$ {indicates the thermal
velocity}, {$\nu$ is the attempt-to-escape frequency},
{
{$E_{i}=i\Delta E$ is the $i$-th energy level
below (or above) a mobility edge},} $g\left(E_{i}\right)$ refers
to the DOS at the $i$-th localized state, and $\delta(t)$
and $\delta(x)$ are Dirac delta functions that establish the initial
conditions for the ToF experiment. These equations can be resolved
via Laplace transforms. The Laplace domain solution for $\hat{p}(x,s)$
is provided by (see to \cite{naito1994,naito1996,nagase1998,ogawa2000}
for further details):
\begin{equation}
\hat{p}(x,s)=\frac{p_{0}}{\mu_{0}F}\exp\left(-\frac{a(s)t_{0}x}{L}\right),\label{eq:3}
\end{equation}
where:
\begin{equation}
a(s)=s\left[1+\sum_{i}\frac{\omega_{i}}{s+g_{i}}\right]=s\left[1+\int_{0}^{E_{\mathrm{mid}}}\frac{\sigma_{p}\nu_{\text{th}}g(E)}{s+\nu\exp(-E/k_{\text{B}}T)}\rd E\right],
\label{eq:4}
\end{equation}
with $t_{0}=\frac{L}{\mu_{0}F}$ and $E_{\text{mid }}$ representing
the midgap energy and $L$ being the sample thickness. Differentiating
$a(s)$ with respect to $\ln(s)$ gives:
\begin{equation}
\frac{\rd a(s)}{\rd\ln(s)}=s\left[1+\int_{0}^{E_{\text{mid }}}\sigma_{p}\nu_{\text{th}}g(E)h(s,E)\rd E\right],
\label{eq:5}
\end{equation}
with
\begin{equation}
h(s,E)=\frac{\nu\exp(-E/k_{\text{B}}T)}{[s+\nu\exp(-E/k_{\text{B}}T)]^{2}}.
\label{eq:6}
\end{equation}
This function $h(s,E)$ has a maximum value at $E_{0}$ and can be
approximated by a delta function:
\begin{equation}
h(s,E)\approx\frac{k_{\text{B}}T}{s}\delta\left(E-E_{0}\right),\label{eq:7}
\end{equation}
where:
\begin{equation}
E_{0}=k_{\text{B}}T\ln\left(\frac{\nu}{s}\right).\label{eq:8}
\end{equation}
For the ToF experiment, the photocurrent $I(t)$ is given by:
\begin{equation}
I(t)=\frac{q\mu_{0}F}{L}\int_{0}^{L}p(x,t)\rd x
\label{eq:9}
\end{equation}
which transforms into the Laplace domain as:
\begin{equation}
\hat{I}(s)=I\left(0\right)\frac{\left[1-\exp\left(-a(s)t_{0}\right)\right]}{a(s)},\label{eq:10}
\end{equation}
with $I\left(0\right)=\frac{q\mu_{0}n_{0}F}{L}$.

Differentiating $\hat{I}(s)$ with respect to $\ln(s)$, we obtain:
\begin{equation}
\frac{\rd}{\rd\ln(s)}\left(\frac{1}{\hat{I}(s)}\right)=\frac{L}{qn_{0}\mu_{0}F}\frac{\rd a(s)}{\rd\ln(s)}B(s),
\label{eq:11}
\end{equation}
where:
\begin{equation}
B(s)=\frac{1-\left[1+a(s)t_{0}\right]\exp\left[-a(s)t_{0}\right]}{\left[1-\exp\left(-a(s)t_{0}\right)\right]^{2}}.
\label{eq:12}
\end{equation}
The localized-state distribution can then be extracted using:
\begin{equation}
g\left(E_{0}\right)=\frac{1}{\sigma_{p}\nu_{\text{th}}k_{\text{B}}T}\left[\frac{1}{B(s)}\frac{\rd}{\rd\ln(s)}\left(\frac{I(0)}{\hat{I}(s)}\right)-s\right],
\label{eq:13}
\end{equation}
where $E_{0}(s)$ is defined by $E_{0}=k_{\text{B}}T\ln(\nu/s)$. 

Equation (\ref{eq:13}) provides the theoretical foundation for extracting
the DOS in amorphous semiconductors from experimental
ToF data. This study applies the Laplace-based formulation
to analyze ToF photocurrent data for pure amorphous selenium (a-Se),
as reported by Benkhedir et al. \cite{Benkhedir2006,Emelianova2006}.

While various techniques exist for estimating the DOS in amorphous
semiconductors, the present approach offers a direct, temperature-resolved
reconstruction of the conduction band DOS without relying on predefined
functional forms or iterative fitting procedures. This provides a
complementary perspective to earlier studies and contributes to addressing
the scarcity of quantitative DOS data specific to pure a-Se.

\subsection{Results and discussion}

In order to construct a current numerically, we proceeded with the
following steps: as well-known, the inverse Laplace of a function
is given by:
\begin{equation}
I(t)=\frac{1}{2\piup \ri}\int_{c-\ri\infty}^{c+\ri\infty}I(s)\re^{st}\rd t,
\label{eq:14}
\end{equation}
which $I(s)$ is defined by equation~(\ref{eq:10}). 

When we use the following change of variables, $z=st$, and $\rd z=s \rd t$,
the integral becomes 
\begin{equation}
I(t)=\frac{1}{2\piup \ri t}\int_{c^{\prime}-\ri\infty}^{c^{\prime}+\ri\infty}I\left(\frac{z}{t}\right)\re^{z}\rd z.
\label{eq:15}
\end{equation}
To compute the final integral, we employ the methodology outlined
in  \cite{serdouk2015,serdouk2020,serdouk2023,goutal2024}. According
to (\ref{eq:15}), the calculated current can be derived as follows:
(i) First, we articulate the function $\re^{z}$ using its Pad\'{e}
approximation expansion; the Pad\'{e} approximation of a function
resembles a Taylor series, but the expansion takes the form of a ratio
of two polynomials. (ii) Secondly, we utilize the residue theorem
to derive the desired function, namely, the calculated current $I(t)$.
Consequently, we computed the inverse Laplace transform numerically,
employing the Pad\'{e} approximation, which utilizes rational functions
with polynomials of eighth degree. {The present work
is grounded in the use of the non-monotonous DOS model
proposed in \cite{serdouk2015} as a basis for investigating
charge transport properties in stabilized amorphous selenium. The
primary objective of this research axis is to evaluate the hole drift
mobility by implementing our custom model of localized state distribution.
Particular attention is given to amorphous selenium doped with 0.2\%
and 0.5\%, arsenic a composition known to improve thermal stability
and widely used in photoconductive materials. To initiate this analysis,
the physical quantity utilized in the computation was $\sigma\nu_{\text{th}}=10^{-7}~\text{cm}^{3}\,\mathrm{s}^{-1}$,
$\nu=10^{12}~\mathrm{s}^{-1}$ and $k_{\text{B}}=8.625\times10^{-5}~\ensuremath{\mathrm{eV\,K}^{-1}}$
is the Boltzmann constant \cite{naito1996}. This simulation serves
as a preliminary step toward validating our proposed DOS
model.}

\begin{figure}[!t]
\begin{centering}
\includegraphics[scale=0.25]{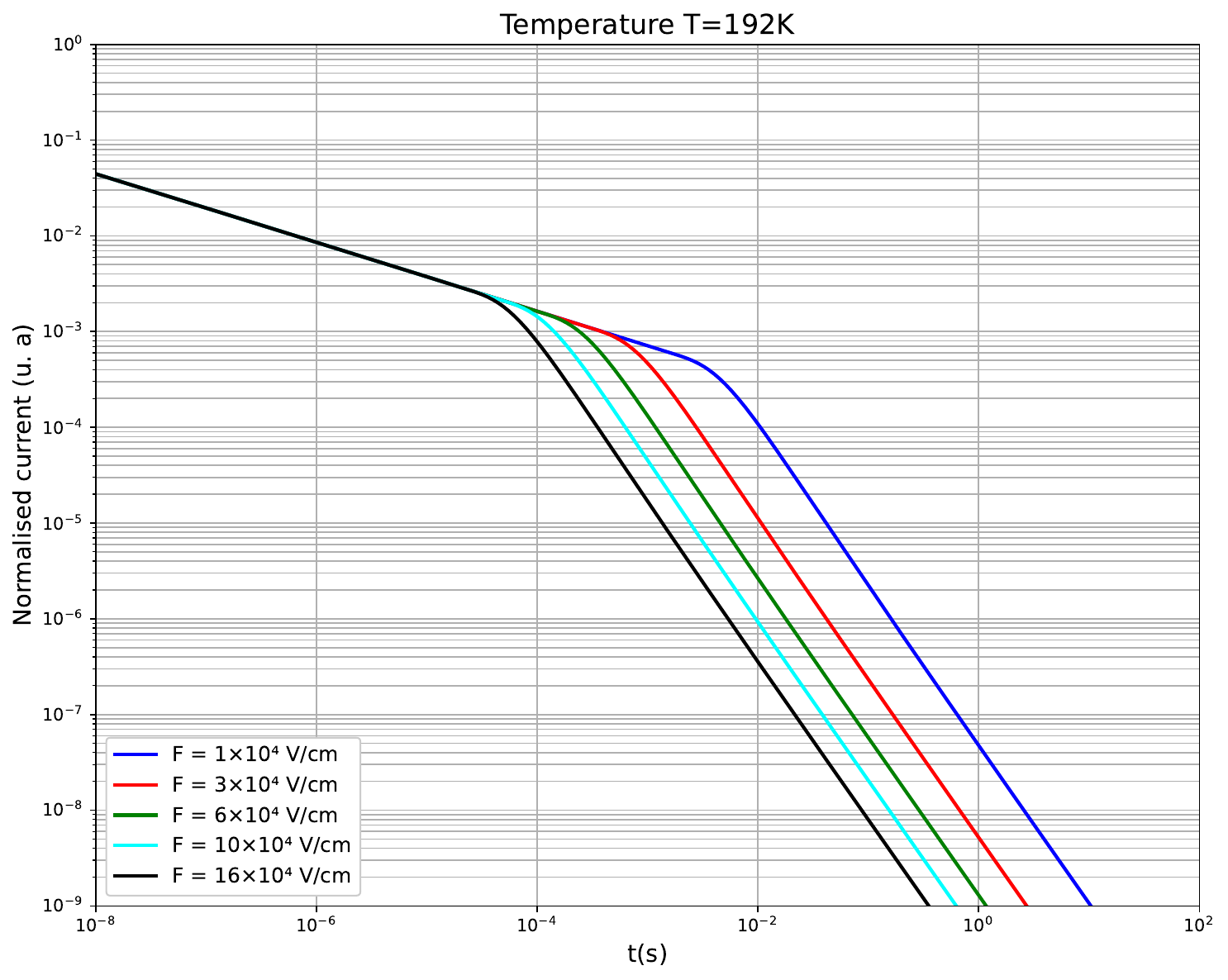}~\includegraphics[scale=0.25]{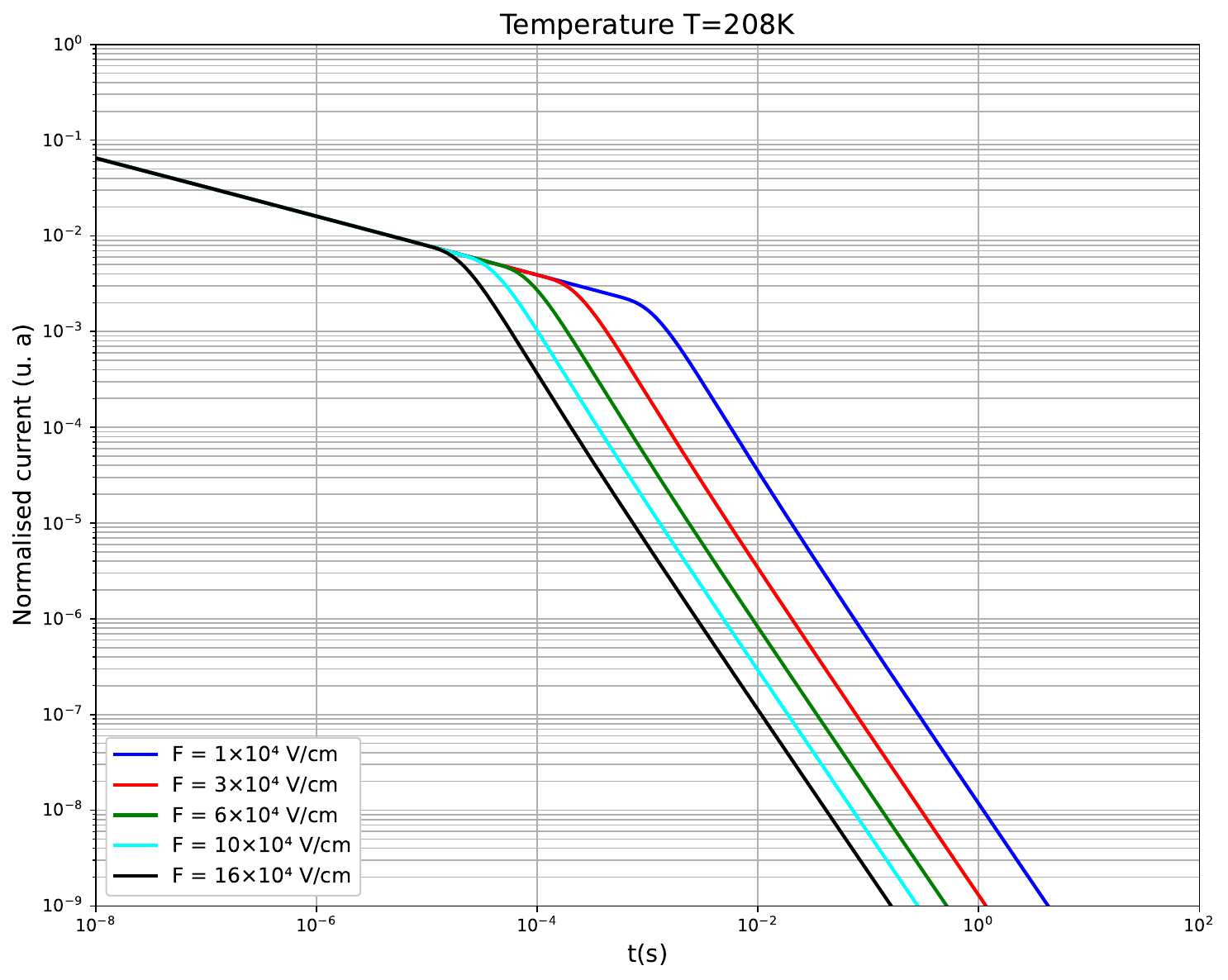}
\par\end{centering}
\begin{centering}
\includegraphics[scale=0.25]{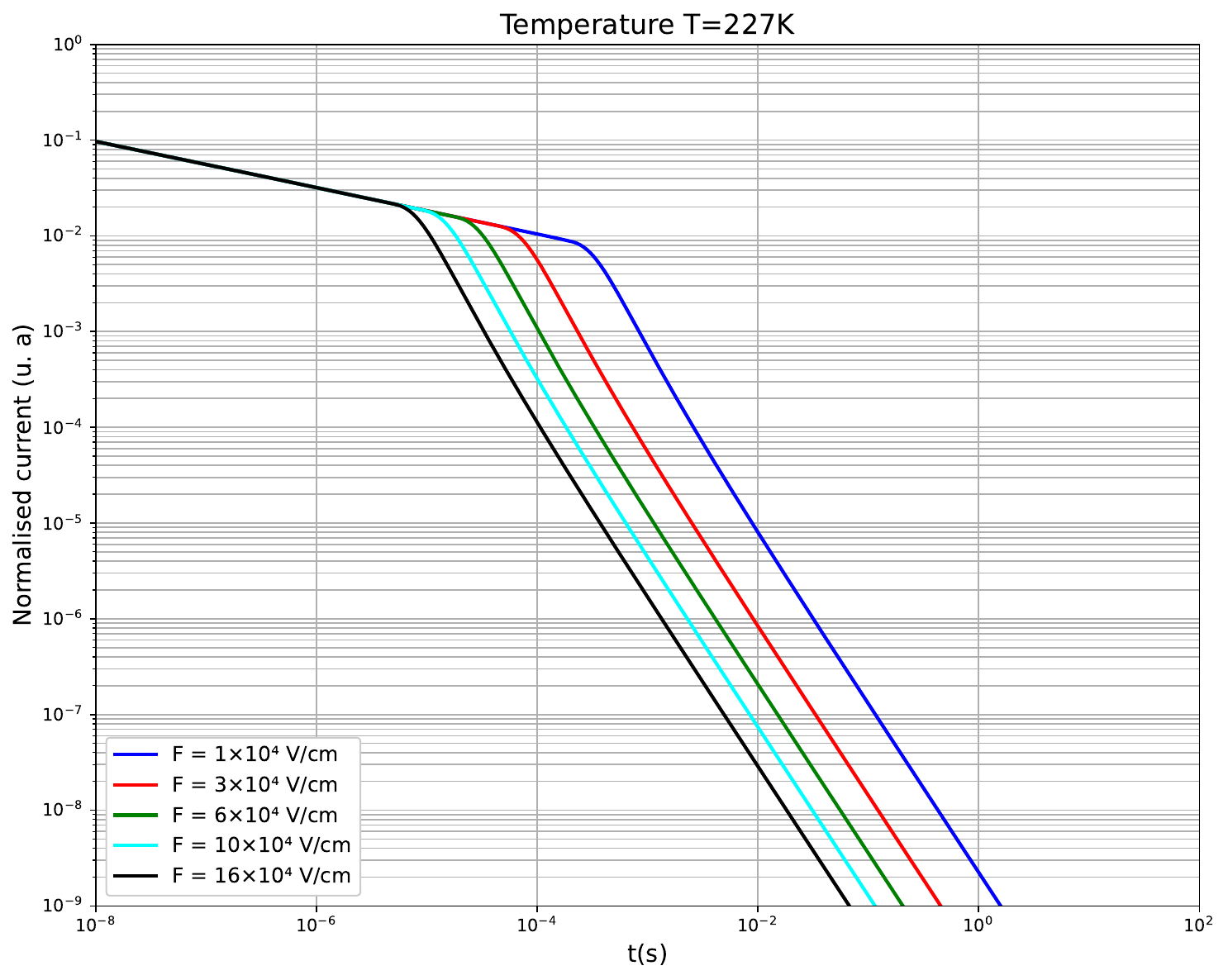}~\includegraphics[scale=0.25]{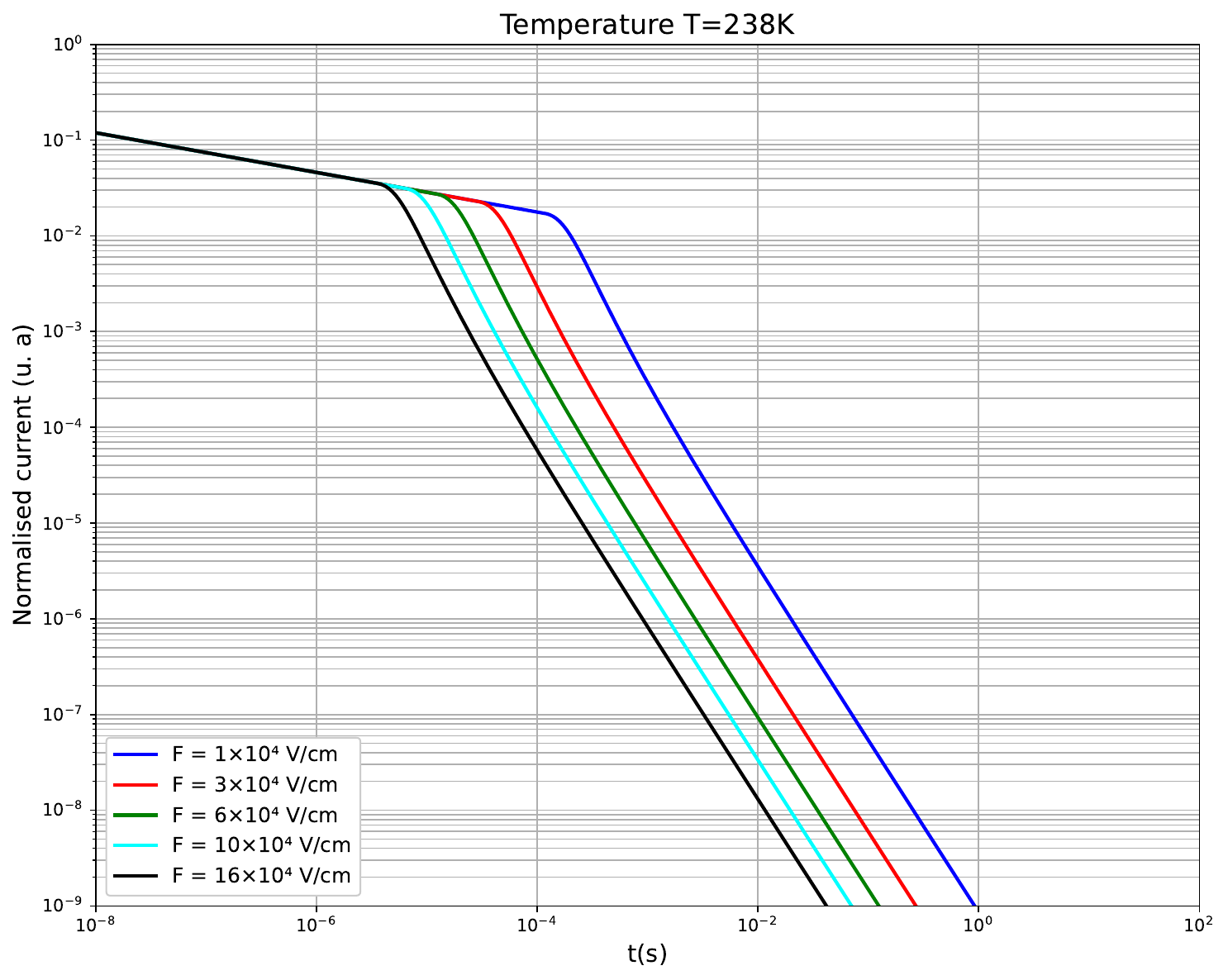}
\par\end{centering}
\caption{(Colour online) Simulated ToF current for the holes in amorphous semiconductors under
varying temperatures and electric fields.}\label{fig:1}
\end{figure}

We are now ready to present and evaluate all the results produced
by the Python software used to simulate our currents. Our algorithm
was implemented in Python~\cite{vanrossum2009}.
Figure~\ref{fig:1} illustrates the normalized photocurrents over
time on a log-log scale for different applied electric fields. For
each value of $F$, four unique temperatures were examined: $T=192$, 208, 227, and $238$~K.
The illustration distinctly indicates that the transient photocurrent
diminishes with time. {The} transit time is determined
at the intersection of the transient photocurrent curve with the time
axis \cite{noolandi1977}, which is essential for calculating the mobility
as a function of the electric field, a subject that will be discussed
in the subsequent section.

Furthermore, as depicted in figure~\ref{fig:1}, all the curves exhibit
analogous behavior in both the pre- and post-transient phases. The
image illustrates the ToF transient photocurrents
under various applied electric fields, indicating that the transit
time is significantly less than the monomolecular recombination lifetime.
As the electric field intensifies, the inflection points migrate to
shorter time intervals, signifying that these points are associated
with the charge-carrier transit time. This facilitates an experimental
differentiation: if the inflection point shifts to shorter durations
with an increase in the electric field, it can be ascribed to the
charge-carrier transit time.

Consequently, drift mobilities can be precisely ascertained, given
that the charge-carrier transit time is considerably less than the
monomolecular recombination lifetime. This criterion is applicable
to non-dispersive transport and is independent of specific physical
factors, including $g(E)$ and $\mu_{0}$.

\subsection{Determination of localized-state distribution}

{We aim at demonstrating that the distributions of localized
states can be ascertained through the analysis of transient photocurrent.
Equation~(\ref{eq:5}) is a Fredholm integral equation of the first kind,
which may arise from an ill-conditioned problem. The derived density
of states (DOS) satisfies all the conditions required to define a
generalized solution to such an ill-posed problem and, therefore,
necessitates a dedicated resolution method.}

{In this section, the adopted approach consists in
approximating the integration kernel of the equation using a suitably
weighted Dirac function. This leads to a simplified expression analogous
to equation~(\ref{eq:13}).}

{To assess the effectiveness of the proposed method,
we computed the transient current response $I(t)$ for a representative
distribution of traps using our numerical procedure. A Laplace transform
was then applied to $I(t)$ to obtain $I(s)$, as defined in equation
(\ref{eq:10}), and the DOS was subsequently calculated using equation~(\ref{eq:13}).}

\begin{figure}[!t]
\begin{centering}
\includegraphics[scale=0.5]{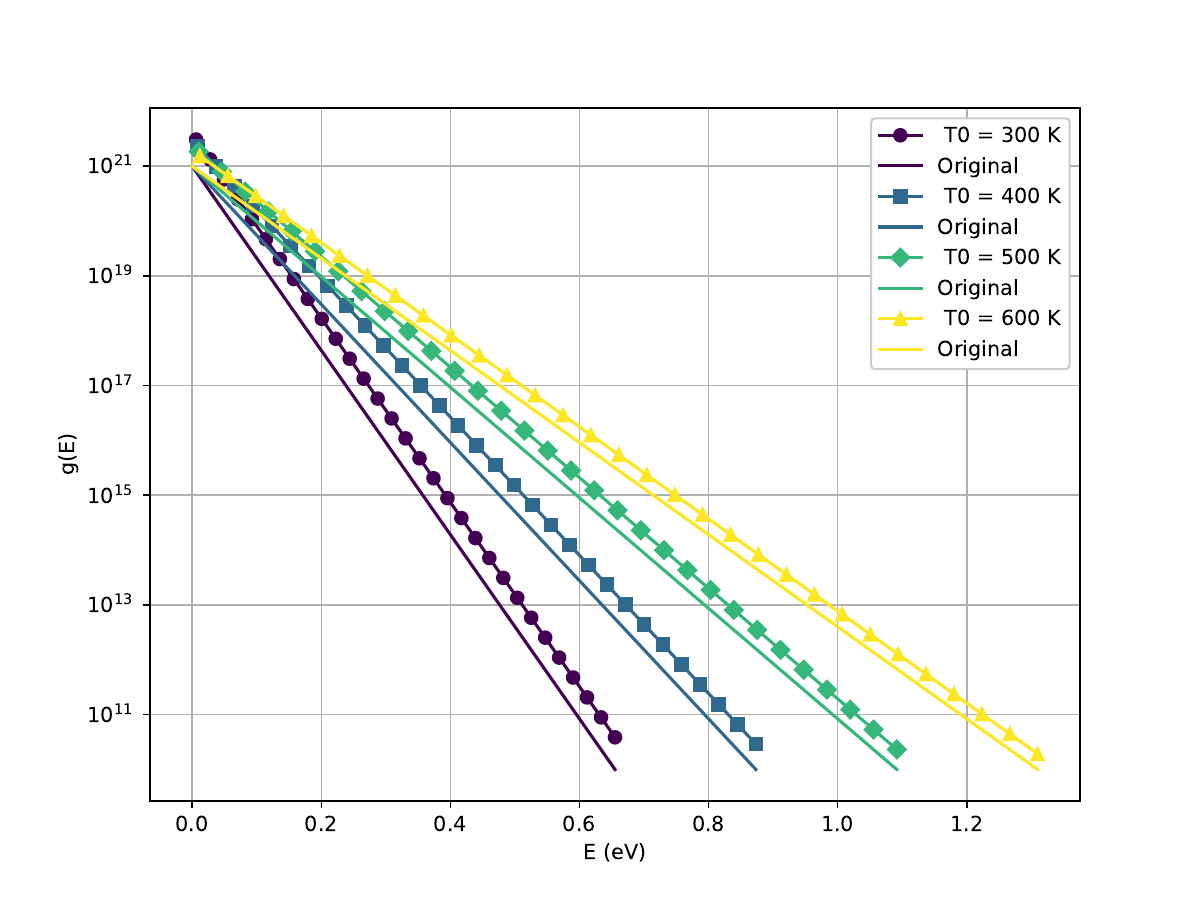}
\par\end{centering}
\caption{(Colour online) Localized-state distributions determined from the numerically calculated
ToF transient photocurrent. The solid lines are original distributions
at $T=300$~K for numerical calculation of the transient photocurrent
using equation~(\ref{eq:13}).}
\label{fig:2}
\end{figure}

{Figure~\ref{fig:2} illustrates the validation process
for selecting an appropriate DOS and reconstructing
its distribution using our numerical methodology. The localized-state
distributions are accurately replicated for $T_{0}>T$, where the
solid lines represent the original DOS at $T=300\,\text{K}$ used
to simulate the transient photocurrents. A strong agreement is observed
between the original distributions and those reconstructed using equation
(\ref{eq:13}), thereby confirming the reliability and robustness of the proposed
Laplace-transform-based method. This result validates our ability
to extract the DOS in pure a-Se and provides a solid foundation for
investigating deeper charge transport mechanisms in amorphous semiconductors.}

{Building on this framework, we conducted an extensive
investigation using the  ToF technique --- widely recognized
for its capability to probe the dynamics of charge carriers in disordered
materials. ToF measurements were performed over a wide range of temperatures
(from $-40^\circ$C to $+23^\circ$C) to analyze the distribution of trap states
within the conduction band tail of pure amorphous selenium.}

\begin{figure}[!t]
\begin{centering}
\includegraphics[scale=0.4]{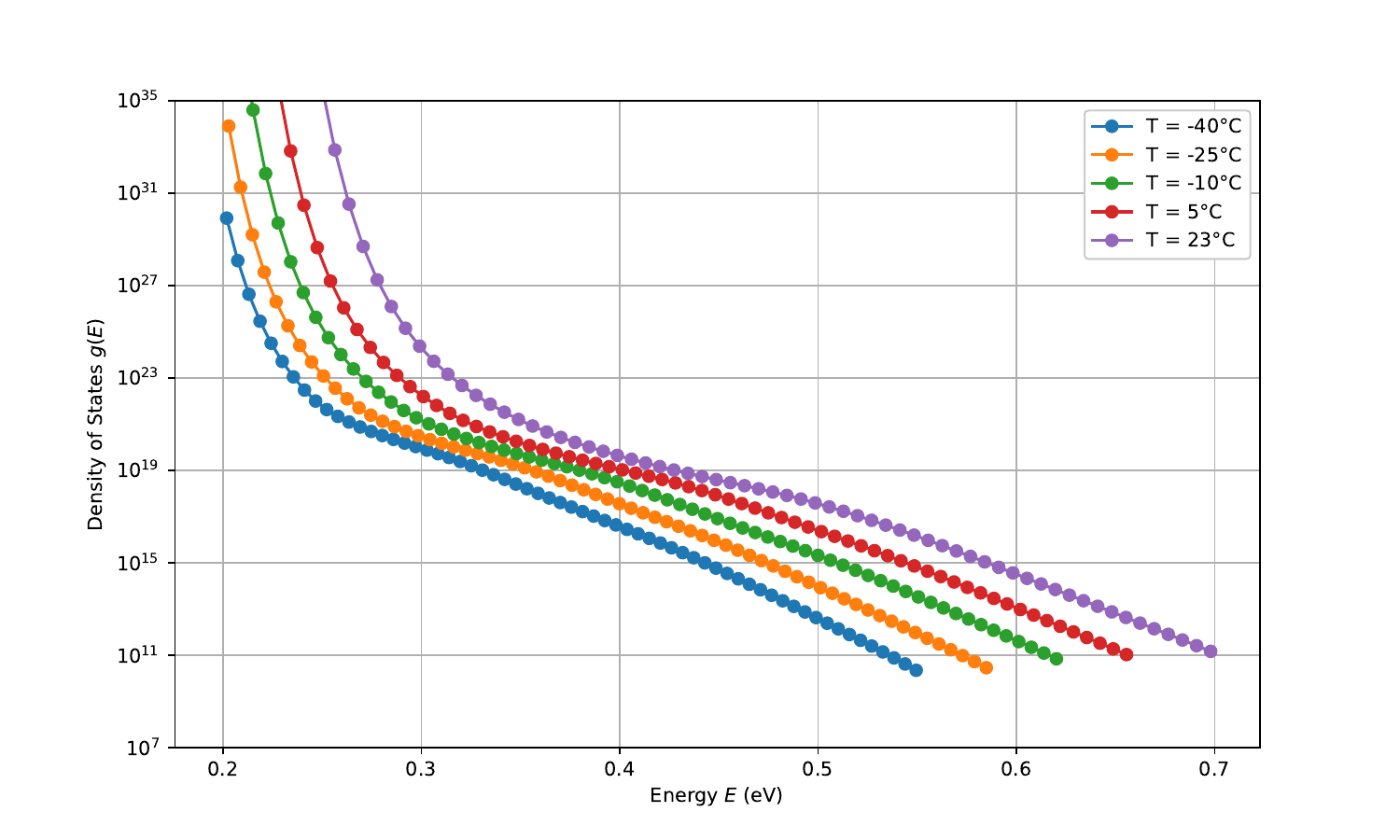}
\par\end{centering}
\caption{(Colour online) Localized-state distributions of the pure a-Se determined from the
numerically calculated ToF transient photocurrent \cite{Benkhedir2006,Emelianova2006}}\label{fig:3}
\end{figure}

{Figure~\ref{fig:3} presents the reconstructed localized-state
distributions derived from ToF transient photocurrent data. These
profiles deviate from a simple exponential decay and instead exhibit
a Gaussian-like shape. Two distinct features are observed:}

{• A shallow trap peak appearing in the range $E-E_{c}\approx0.28-0.30\,\text{eV}$,
which is clearly visible at low temperatures ($-40^\circ$C, $-25^\circ$C).
At these temperatures, reduced thermal energy limits the carrier access
to states near the conduction band edge, enhancing the visibility
of shallow traps.}

{• A deeper trap peak centered around $E-E_{c}\approx0.45-0.50\,\text{eV}$,
which becomes increasingly prominent at higher temperatures (from
$-10^\circ$C to $+23^\circ$C), as carriers gain sufficient thermal energy
to populate and escape the deeper localized states.}

{Furthermore, the position of the shallow trap level,
identified in the present study near $0.30\,\text{eV}$ below the
conduction band mobility edge, is in excellent agreement with the values
derived from pre-transit photocurrent analyses. Specifically, earlier
studies have reported a shallow defect state energy of $E_{\text{se}}=0.28\pm0.02\,\text{eV}$,
extracted from the early-time regime of ToF measurements~\cite{benkhedir2008}.
This state reflects a population of shallow traps that govern the carrier
behaviour before transit. The excellent agreement with our reconstructed
DOS confirms the accuracy of the Laplace-transform method in resolving
the energy distributions near the conduction edge.}

{These temperature-dependent variations provide crucial
insight into the thermal activation dynamics of carriers and confirm
the presence of energetically distinct trapping centers. The position
and nature of these two peaks are in excellent agreement with earlier
findings, which revealed similar defect states near $0.30$~eV and $0.45-0.50$~eV
in a-Se through numerical simulations of ToF photocurrents \cite{Koughia2005}.
However, unlike their study, our broader temperature range and time
resolution allow for a more accurate resolution of these defect states,
and do not indicate the presence of additional deep states below 0.6\,eV.
The discrepancy probably arises from the experimental limitations in
their work --- namely, a narrow time window centered around the transit
time and measurements performed exclusively at room temperature --- conditions
that are insufficient to isolate contributions from deeper levels.}

{Importantly, the results of the present study reveal
a remarkable symmetry in the distribution of localized states between
the valence and conduction band tails in amorphous selenium. The reconstructed
DOS, extracted via Laplace-transform analysis from ToF data, displays
comparable energetic spreads and defect densities on both sides of
the mobility gap. This symmetry strongly suggests that analogous trapping
and release mechanisms govern the transport of both electrons and
holes, reflecting the intrinsic structural disorder of the amorphous
matrix.}

{This observation directly complements our earlier
work \cite{serdouk2015}, in which we focused on the valence band
side of pure a-Se. In that study, we applied the same Laplace-transform
technique to transient photoconductivity (TPC) measurements and revealed
a non-monotonous DOS profile with two discrete defect levels:}

{• A shallow level approximately 0.30\,eV above the
valence band edge ($E_{v}$), associated with local structural distortions
such as dihedral angle variations.}

{• A deeper level near 0.45\,eV, attributed to the
presence of $D^{-}$ centers, in agreement with the negative-$U$ model
\cite{benkhedir2004,benkhedir2008}.}

{As in the conduction side study, the shallow level
in the valence band is more prominent at low temperatures, while the
deeper level becomes significant at elevated temperatures --- demonstrating
a consistent temperature dependence of trap activation across both
band edges.}

{Furthermore, we extended this investigation to arsenic-doped
a-Se. In this context, shallow defects disappeared and deeper levels
became dominant. This behavior was attributed to arsenic atoms having
modified the local bonding environment, which led to a smoother distribution
of localized states in the valence band tail. Using a least-squares
fitting procedure, we reconstructed the DOS profiles for samples doped
with 0.2\% and 0.5\% As, achieving high accuracy in replicating
experimental photocurrents.}

{Taken together, these studies --- both performed by
the present authors --- provide a comprehensive insight into the energetic
structure of the localized states in a-Se. The symmetric and structured
distribution of states on both sides of the mobillity gap demonstrates
that trapping phenomena are energetically balanced for electrons and
holes. This balance has significant implications for the modelling
of carrier transport in amorphous semiconductors and supports the
use of unified descriptions of localized states across both the conduction
and valence bands.}


An essential aspect of analyzing the transport characteristics in
amorphous selenium is the precise assessment of carrier drift mobility
and lifetime $\tau$. Multiple studies have emphasized the scientific
and technological significance of these factors \cite{naito1994,naito1996}.
{To deepen this understanding and support our experimental
findings, we implemented numerical simulations of hole transport in
stabilized amorphous selenium. The simulations aimed at reproducing
and interpreting photocurrent transients, allowing for a detailed investigation
of the temperature and field dependence of hole mobility.}

\section{Simulation of hole mobility in amorphous semi-conductors}

We additionally corroborated the non-monotonous model of the localized
state density distribution using the simulation of hole photocurrents.
In this instance, we sought to resolve the continuity equations under
various scenarios. The first objective was to replicate transient
 ToF photocurrents utilizing the identical functions
for the localized state density distribution as specified in the preceeding
equations.

Our inverse Laplace software enabled us to acquire currents for various
temperatures and fields. The aim is to assess the hole mobility in
{stabilized} a-Se. This section employs a widely recognized
method from the literature, which involves calculating the transit
time from the simulated ToF currents at a specified temperature and
field utilized in prior experimental investigations. The hole mobility
$\mu_{h}$ can be derived using the subsequent equation \cite{gill1972,marshall1972,kasap1985}:
\begin{equation}
\mu_{h}=\frac{L}{t_{\text{r}}F},\label{eq:16}
\end{equation}
where $L$ denotes the sample thickness and $t_{\text{r}}$ signifies the transit
time. {The transit time was determined using the tangent
intersection method, by identifying the point where linear fits to
the pre-transit and post-transit regions of the ToF current curve
intersect.}

All computations in this study were executed in this manner. The simulated
mobility values for the two samples under varying field and temperature
conditions are presented in table~\ref{tab:1} . The correlation
between mobility and $T^{-1}$ is accurately represented by a linear
regression.

\begin{table}
\caption{Simulated hole mobility in {stabilized} a-Se.}
\label{tab:1}
	\vspace{0.2cm}
\centering{}%
\begin{tabular}{|c|c|c|c|c|c|}
\hline 
$\mathrm{F}(\mathrm{V}/\mathrm{cm})$ & $10^{4}$ & $3\times10^{4}$ & $6\times10^{4}$ & $10^{5}$ & $1.6\times10^{5}$\tabularnewline
\hline 
$\mu_{h}(T=192$K$)\left(\mathrm{cm}^{2}V^{-1}s^{-1}\right)$ & 0.00125 & 0.001667 & 0.002381 & 0.003165 & 0.004085\tabularnewline
$\mu_{h}(T=208$K$)\left(\mathrm{cm}^{2}V^{-1}s^{-1}\right)$ & 0.003876 & 0.00463 & 0.005952 & 0.007937 & 0.012\tabularnewline
$\mu_{h}(T=227$K$)\left(\mathrm{cm}^{2}V^{-1}s^{-1}\right)$ & 0.013 & 0.016 & 0.017 & 0.02 & 0.024\tabularnewline
$\mu_{h}(T=238$K$)\left(\mathrm{cm}^{2}V^{-1}s^{-1}\right)$ & 0.024 & 0.025 & 0.028 & 0.031 & 0.037\tabularnewline
\hline 
\end{tabular}
\end{table}

\begin{figure}[!t]
\begin{centering}
\includegraphics[scale=0.5]{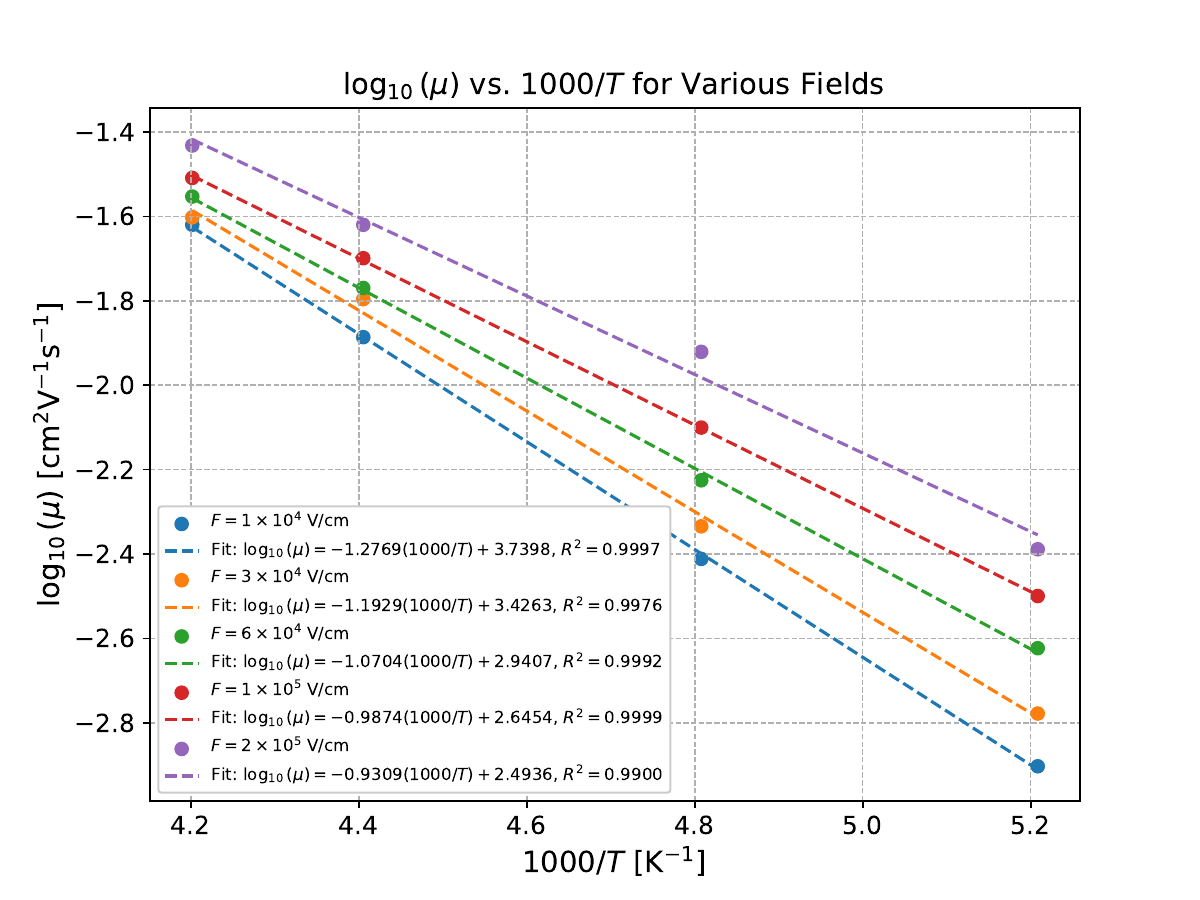}
\par\end{centering}
\caption{(Colour online) Logarithm of hole mobility as a function of inverse temperature for
various electric fields.}\label{fig:4}
\end{figure}

We utilize Arrhenius plots to ascertain the activation energy. The
Arrhenius plots of mobility clearly demonstrate that temperature substantially
influences the values of $\mu$. The Arrhenius plot is a prevalent
analytical instrument in semiconductor physics, utilized to investigate
the temperature dependence of charge carrier mobility. This graphic
depicts the logarithm of mobility $(\log\mu)$ against the inverse
of temperature $(1/T)$. It is extensively utilized to calculate activation
energies and comprehend the temperature-dependent characteristics
of charge transport.

The correlation between mobility $(\mu)$ and temperature $(T)$ often
adheres to an exponential model.
\begin{equation}
\mu=\mu_{0}\exp\left(-\frac{E_{a}}{k_{\text{B}}T}\right),
\label{eq:17}
\end{equation}
where: $\mu_{0}$ denotes the pre-exponential factor (mobility at
infinite temperature) and $E_{a}$ represents the activation energy. 

Applying the logarithm to both sides results in:
\begin{equation}
\log\mu=\log\mu_{0}-\frac{E_{a}}{k_{\text{B}}T}.
\label{eq:18}
\end{equation}
The slope of the Arrhenius plot$~(\log\mu\,\text{vs}\,1/T)$ in this
linearized format can be utilized to ascertain the activation energy
$E_{a}$. The activation energy denotes the thermal energy required
for charge carriers to surmount obstacles like traps or flaws inside
the material. The Arrhenius figure elucidates the relationship between
mobility and temperature variations. In numerous amorphous semiconductors,
such as a-Se (amorphous selenium), mobility generally rises with temperature,
signifying a thermally triggered transport mechanism. Essential points
to consider:

i)~A steeper slope on the Arrhenius plot indicates a greater activation
energy, signifying more pronounced trapping or scattering effects. 

ii)~A smaller slope suggests reduced activation energy, potentially signifying
fewer traps or more extensive states that facilitate improved carrier
delocalization. 

In amorphous semiconductors such as a-Se, the Arrhenius plot is commonly
utilized to investigate the hole mobility. It can furnish critical insights
regarding localized states, the DOS, and the effects
of doping (e.g., arsenic doping). Fluctuations in the slope or departures
from a linear trajectory may signify shifts between distinct charge
transport regimes, such as from non-dispersive to dispersive transport.
The Arrhenius plot of mobility is an effective instrument for examining the
charge transport pathways in disordered materials. It facilitates
the extraction of essential data, such as activation energy, and clarifies
the influence of temperature on mobility, providing enhanced understanding
of the electrical structure of amorphous semiconductors.

Return to our case now. {To investigate the temperature
dependence of hole mobility, we plot $\log\textgreek{\textmu}$ as a function
of the inverse temperature (1/T) for each applied electric field,
as shown in figure~\ref{fig:4}. This representation reveals a temperature-dependent
variation in mobility and provides a clearer interpretation of the
thermally activated transport mechanism.}

{The temperature dependence of hole mobility as a function
of the inverse temperature $1/T$ is well approximated by a linear
relationship. Arrhenius plots clearly demonstrate that temperature
has a significant influence on mobility. Within the temperature range
of $192$ K to $238$ K, mobility increases with temperature, in excellent
agreement with previously reported results \cite{Fogal2014}. From
the linear regressions of the $\log\mu\upsilon$
versus $1/T$
curves, activation energies were extracted from the slopes of the
linear regions. Additionally, variations in these slopes under different
electric field strengths highlight the field dependence of the activation
energy. The extracted values are summarized in table~\ref{tab:2}.
 The high correlation coefficients confirm that the hole mobility in both
studied samples exhibits an exponential dependence on temperature.
Moreover, the analysis of the mobility variation with respect to electric
field at different temperatures reveals that the activation energies
are field-dependent. This behaviour is indicative of the Poole-Frenkel
effect, which describes the field-assisted thermal emission of charge
carriers from localized trap states.}

{The Poole-Frenkel coefficient $\beta_{\text{PF}}$ is a
parameter that characterizes the field-induced reduction of the potential
barrier for charge carriers confined in localized states within insulating
or semiconducting materials \cite{hill1971}. It provides an insight
into the defect structure of the material and the mechanisms governing the
charge transport by quantifying how an electric field facilitates
the release of the trapped carriers. The effect occurs when an applied
electric field lowers the energy barrier associated with a Coulombic
trap --- a localized defect that attracts a charge --- thereby increasing
the probability of carrier escape and enhancing electrical conduction.}
\begin{center}
\begin{table}
\caption{Activation energy values.}
\label{tab:2}
\vspace{0.2cm}
\centering{}%
\begin{tabular}{|c|c|c|c|c|c|}
\hline 
$\mathrm{V}($ volt $)$ & 50 & 150 & 300 & 500 & 800\tabularnewline
\hline 
\hline 
$F(\mathrm{~V}/\mathrm{cm})$ & $10^{4}$ & $3\times10^{4}$ & $6\times10^{4}$ & $10\times10^{4}$ & $16\times10^{4}$\tabularnewline
\hline 
$E_{h}(\mathrm{eV})$ & $0.254$ & $0.237$ & $0.213$ & $0.196$ & $0.185$\tabularnewline
\hline 
\end{tabular}
\end{table}
\par\end{center}

\begin{figure}
\begin{centering}
\includegraphics[scale=0.4]{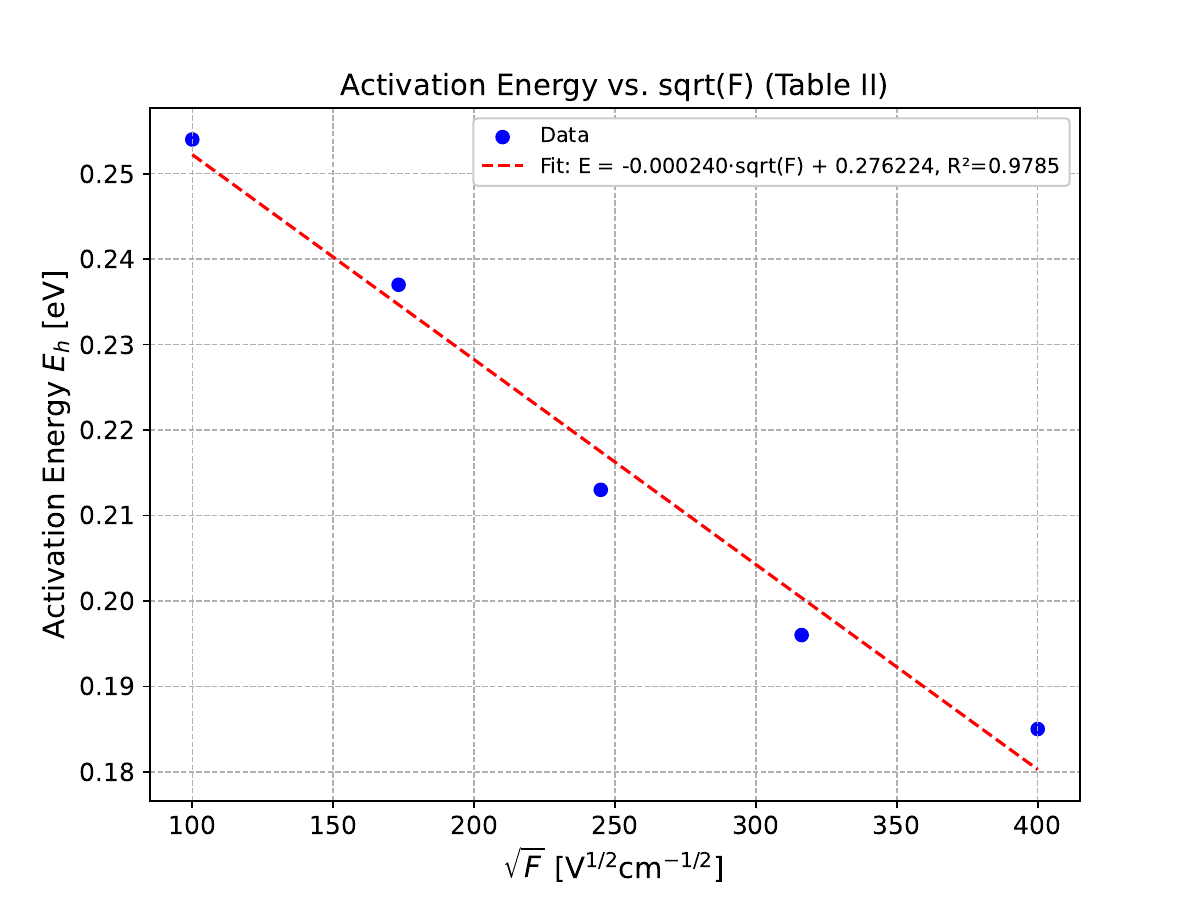}
\par\end{centering}
\caption{(Colour online) Variation of activation energy as a function of the electric field.}
\label{fig:5}
\end{figure}

{The Poole-Frenkel effect is typically expressed
by the subsequent relationship:}
\begin{equation}
\mu=\mu_{0}\exp\left(-\frac{E_{a}-\beta_{\mathrm{PF}}\sqrt{F}}{k_{\text{B}}T}\right),
\label{eq:19}
\end{equation}
where: $\mu$ represents mobility, $\mu_{0}$ denotes the pre-exponential
factor, $E_{a}$ signifies the activation energy at zero field, and
$\beta_{\mathrm{PF}}$ indicates the Poole-Frenkel coefficient. The
Poole-Frenkel coefficient is expressed as follows:
\begin{equation}
\beta_{\mathrm{PF}}=\sqrt{\frac{q^{3}}{\piup\varepsilon_{0}\varepsilon_{r}}},
\label{eq:20}
\end{equation}
where $q$ denotes the elementary charge $(1.6\times10^{-19}\,\text{C})$, $\varepsilon_{0}$
represents the vacuum permittivity ($8.854\times10^{-12}\,F/m$), and
$\varepsilon_{r}$ signifies the relative permittivity of the material.

{In our results, the activation energy decreases from
$0.254$~eV to $0.185$~eV. When these values are fitted to the
Poole-Frenkel relation, a linear dependence on $F^{1/2}$ is observed,
as shown in figure~\ref{fig:5}. The slope of the linear fit yields
a Poole-Frenkel coefficient of $\beta_{\text{PF}}=2.4\times10^{-4}\,\text{eV}\,(\text{V}\,\text{cm}^{-1})^{-1/2},$
and the extrapolated activation energy at zero field is approximately
$0.28$~eV. These results provide strong evidence that charge transport
in the studied material is dominated by a field-assisted thermal emission
mechanism.}

{It is physically significant to note that similar
field-dependent behaviors of activation energy have been reported
in prior studies. In the case of undoped amorphous selenium, an activation
energy of $0.28\pm0.02\,\text{eV}$ was observed at temperatures below
$250$~K \cite{marshall1972}, indicating a thermally activated transport
regime. For stabilized (arsenic-doped) amorphous selenium, the reported
activation energy was approximately $0.21$~eV~\cite{Fogal2014},
which is in close agreement with the value extracted in the present
work. Moreover, the Poole-Frenkel coefficient $\beta_{\text{PF}}$
reported in \cite{Benkhedir2010} estimated as $\beta_{\text{PF}}=2.7\times10^{-4}\,\text{eV}\,(\text{V}/\text{cm})^{-1/2}$
is in good agreement with the value deduced in this study and corresponds
to the doped case. This consistency reinforces the interpretation
that hole transport occurs via field-assisted thermal emission from
localized trap states. The same study \cite{Benkhedir2010} also supports
the presence of a non-monotonous density of localized states near the
valence band edge, further validating our theoretical approach.}

\section{{Conclusion}}

{In this study, we performed a comprehensive numerical
investigation of charge transport in amorphous selenium (a-Se) based
on a Laplace-transform analysis of  ToF transient
photocurrents. The reconstructed DOS on the conduction
band side in pure a-Se revealed a symmetric distribution with respect
to the valence band side, as established in our previous studies,
indicating balanced trapping dynamics across the mobility gap.}

{To validate the transport mechanisms, we simulated the
hole mobility using a known DOS model for stabilized a-Se. The simulated
data exhibited an Arrhenius-type temperature dependence, with activation
energies decreasing from approximately $0.25$~eV at low fields to
$0.18$~eV at higher fields. The extracted Poole-Frenkel coefficient,
$\beta_{\text{PF}}=2.4\times10^{-4}\,\text{eV}\,(\text{V}/\text{cm})^{-1/2},$
supports the interpretation that the hole transport is governed by field-assisted
thermal emission from localized states.}

{The agreement between our simulation results and established
experimental data confirms the robustness of the proposed model and
method.}

\bibliographystyle{cmpj}
\bibliography{referenceserdouk}

\ukrainianpart

\title{
	Дослідження густини станів та рухливості носіїв заряду в аморфних напівпровідниках за допомогою аналізу фотоструму за часом прольоту (ToF)}
\author{
	Ф. Сердук, 
	А. Бумалі,
	М. Л. Бенхедір,
	І. Гутал
}

\address{Лабораторія теоретичної та прикладної фізики, Університет Ешахіда Шейха Ларбі Тебессі, Тебесса, Алжир}

\makeukrtitle

\begin{abstract}
	\tolerance=3000%
	У даному дослідженні розглядаються характеристики електронного переносу	 аморфних напівпровідників за допомогою вимірювань часу прольоту (ToF) та числового моделювання. Основною метою є визначення густини станів (DOS) в аморфному селені (a-Se) та оцінка залежності рухливості дірок від температури та електричного поля. Комплексне дослідження локалізованих станів у межах щілини рухливості проведено за допомогою перетворення Лапласа для перехідних процесів фотоструму ToF у поєднанні з моделлю множинних пасток. Такий підхід дозволяє точно реконструювати DOS у широкому діапазоні температур, що уможливлює чітку ідентифікацію поверхневих та внутрішніх пасток та виявлення термічно активованих механізмів переносу. Модельовані струми ToF також використовуються для оцінки рухливості дрейфу дірок за різних температурних та польових умов. Енергії активації отримані з графіків Арреніуса для значень рухливості. Результати підтверджують фізично узгоджений опис електронної структури в a-Se а також застосовність методів на основі перетворення Лапласа для вивчення переносу заряду в невпорядкованих напівпровідниках.

	\keywords аморфний селен (a-Se), час прольоту (ToF), перехідний фотострум (TPC), перетворення Лапласа, модель множинного захоплення (MTM), густина станів (DOS)
\end{abstract}

\end{document}